\documentclass[twocolumn]{aastex631}
\newcommand{\Japanese}[1]{\begin{CJK}{UTF8}{ipxm}(#1)\end{CJK}}

\usepackage{appendix}
\usepackage{CJKutf8, color}
\newcommand{\Chinese}[1]{{\begin{CJK*}{UTF8}{gbsn}(#1)\end{CJK*}}}

\usepackage{ulem}
\DeclareRobustCommand{\erase}{\bgroup\markoverwith{\textcolor{black}{\rule[.5ex]{2pt}{0.4pt}}}\ULon}

\DeclareRobustCommand{\adsout}{\bgroup\markoverwith{\textcolor{blue}{\rule[.5ex]{2pt}{0.4pt}}}\ULon}

\DeclareRobustCommand{\hlsout}{\bgroup\markoverwith{\textcolor{red}{\rule[.5ex]{2pt}{0.4pt}}}\ULon}

\begin{document}
\title{The Impacts of Neutron-Star Structure and Base Heating on Type I X-Ray Bursts and Code Comparison}
\correspondingauthor{Helei Liu}
\email{heleiliu@xju.edu.cn}
\correspondingauthor{Akira Dohi}
\email{dohiak@hiroshima.ac.jp}
\author{Guoqing Zhen \Chinese{甄国庆}}
\affiliation{School of Physical Science and Technology, Xinjiang University, Urumqi 830046, China}
\author{Guoliang L\"{u} \Chinese{吕国梁}}
\affiliation{Xinjiang Astronomical Observatory, Chinese Academy of Science, 150 Science 1-Street, Urumqi 830011, China}
\affiliation{School of Physical Science and Technology, Xinjiang University, Urumqi 830046, China}
\author{Helei Liu \Chinese{刘荷蕾}}
\affiliation{School of Physical Science and Technology, Xinjiang University, Urumqi 830046, China}
\author{Akira Dohi \Japanese{土肥明}} 
\affiliation{Department of Physics, Graduate School of Advanced Science and Engineering, Hiroshima University, Higashi-Hiroshima, Hiroshima 739-8526, Japan}
\affiliation{Interdisciplinary Theoretical and Mathematical Sciences Program (iTHEMS), RIKEN, wako, Saitama 351-0198, Japan}
\author{Nobuya Nishimura \Japanese{西村信哉}}
\affiliation{Astrophysical Big Bang Laboratory, Cluster for Pioneering Research, RIKEN, Wako, Saitama 351-0198, Japan}
\affiliation{RIKEN Nishina Center for Accelerator-Based Science, RIKEN, Wako, Saitama 351-0198, Japan}

\author{Chunhua Zhu \Chinese{朱春花}}
\affiliation{School of Physical Science and Technology, Xinjiang University, Urumqi 830046, China}
\author{Liyu Song \Chinese{宋利宇}}
\affiliation{School of Physical Science and Technology, Xinjiang University, Urumqi 830046, China}
\author{Weiyang Wang \Chinese{王维扬}}
\affiliation{Department of Astronomy, Peking University, Beijing 100871, China}
\affiliation{Kavli Institute for Astronomy and Astrophysics, Peking University, Beijing 100871, China}
\author{Renxin Xu \Chinese{徐仁新}}
\affiliation{Department of Astronomy, Peking University, Beijing 100871, China}
\affiliation{Kavli Institute for Astronomy and Astrophysics, Peking University, Beijing 100871, China}
\label{firstpage}

\begin{abstract}

Type I X-ray bursts are rapidly brightening phenomena triggered by thermonuclear burning on accreting layer of a neutron star (NS). The light curves represent the physical properties of NSs and the nuclear reactions on the proton-rich nuclei.
The numerical treatments of the accreting NS and physics of the NS interior are not established, which shows uncertainty in modelling for observed X-ray light curves.
In this study, we investigate theoretical X-ray-burst models, compared with burst light curves with GS~1826-24 observations. We focus on the impacts of the NS mass, the NS radius, and base-heating on the NS surface using the \texttt{MESA} code.
We find a monotonic correlation between the NS mass and the parameters of the light curve. The higher the mass, the longer the recurrence time and the greater the peak luminosity. While the larger the radius, the longer the recurrence time, the peak luminosity remains nearly constant. In the case of increasing base heating, both the recurrence time and peak luminosity decrease. 
We also examine the above results using with a different numerical code, \texttt{HERES}, based on general relativity and consider the central NS.
We find that the burst rate, burst energy and burst strength are almost same in two X-ray burst codes by adjusting the base-heat parameter in MESA (the relative errors $\lesssim5\%$), while the duration time and the rise time are significantly different between (the relative error is possibly $\sim50\%$).
The peak luminosity and the e-folding time are ragged between two codes for different accretion rates.

\end{abstract}

\keywords{X-rays: bursts --- stars: neutron --- X-rays: binaries --- nuclear reactions,abundances}

\section{Introduction}

Type I X-ray bursts are periodic eruptions caused by unstable thermonuclear burning on the surface of neutron stars (NSs) in low-mass X-ray binary (LMXB) systems \citep{Joss1977Natur.270..310J, Parikh2013PrPNP..69..225P}. The NS accreted matter from the companion star, overflowed through the Roche lobe, and formed an envelope on the surface of the neutron star. Under the action of gravity, the accreted matter was continuously compressed and heated, thereby increasing the temperature and density, when the energy generation rate is greater than the cooling rate, thermonuclear unstable combustion will occur, resulting in type I X-ray bursts \citep{Woosley1976Natur.263..101W, Lewin1993SSRv...62..223L, Bildsten2000AIPC..522..359B, Galloway2021ASSL..461..209G}. The accreted matter mainly provides energy for type I X-ray bursts through 3$\rm \alpha$ reaction, CNO cycle and rp-process, etc \citep{Wallace1981ApJS...45..389W, Taam1985ARNPS..35....1T, Bildsten1998ASIC..515..419B, Galloway2008ApJS..179..360G}. Burning produces a heavy accumulation of ash, and as new material continues to pile on top of it, the accreted material undergoes gradual compression until it reaches a condition for ignition, producing another burst sequence.

Since the first discovery of the X-ray burst in 1975 \citep{Belian1976ApJ...206L.135B, Grindlay1976ApJ...205L.127G}, more than 7000 events from 118\footnote{https://personal.sron.nl/$\sim$jeanz/bursterlist.html} bursting sources have been observed so far \citep{Galloway2020ApJS..249...32G}. By comparing with observations, theoretical models can be calibrated and the physical properties of NSs can be constrained. \citep{Cromartie2020NatAs...4...72C}.  GS1826-24 is one of the preferred sources
because of a nearly uniform accretion rate and regular burst behaviour, which is called ``clock burst'' or ``textbook'' burst \citep{Ubertini1999ApJ...514L..27U, Bildsten2000AIPC..522..359B}. X-ray burst models require input parameters regarding the accreted fuel composition (X,Y,Z), mass accretion rate ($\dot{M}$), base heating ($Q_b$), mass ($M$) and radius ($R$) of NS, as well as nuclear reaction rates. \cite{Heger2007ApJ...671L.141H} studied the effect of metallicity ($Z$) and mass accretion rate on the theoretical light curves, by the comparison with the light curve of GS1826-24, they estimated the initial metallicity and the accretion rate of GS1826-24. \cite{Meisel2018ApJ...860..147M} investigated the sensitivity of models to varied accretion rate, base heating, metallicity and the nuclear reaction rate $^{15}\rm O(\alpha,\gamma)^{19}\rm Ne$, by model-observation comparisons, they constrained the shallow heating in GS 1826-24 should be below 0.5 MeV/u. The influence of nuclear reaction rate uncertainties on NS properties also has been studied from X-ray burst model-observation comparisons \citep{Meisel2019ApJ...872...84M}. 

The above X-ray burst simulations are based on KEPLER \citep{Heger2007ApJ...671L.141H} or MESA \citep{Meisel2018ApJ...860..147M,Meisel2019ApJ...872...84M}, which consider the NS envelope using inner boundary conditions with fixed NS mass and radius ($1.4M_{\odot}$ and 11.2 km).
The effects of the NS mass and radius on thermonuclear flashes are investigated by \cite{Joss1980ApJ...238..287J} and \cite{Ayasli1982ApJ...256..637A} using the stellar evolution code ASTRA. They adopted $M=1.4M_{\odot}$ and $R=6.57~\rm km$ as a standard case, and vary mass to $M=0.705M_{\odot}$ as the low mass case and radius to $R=13.14~\rm km$ as the large radius case. The results show that the recurrence time, accumulated mass, burst energy, burst strength and peak luminosity have obvious change. However, the results are not consistent with the recent NS mass--radius constraint \citep{Steiner2010ApJ...722...33S, Abbott2018PhRvL.121p1101A}.

Recently, \cite{Dohi2020PTEP.2020c3E02D, Dohi2021ApJ...923...64D, 2022ApJ...937..124D} studied X-ray bursts using a general relativistic stellar-evolution code with several NS equation of states (EOSs). They focused on the microphysics inside NSs (e.g., the mass and radius with different EOSs and the NS cooling process). By comparing with the burst parameters of GS 1826-24, they constrained the EOS and the NS mass and raidus. Meanwhile,
\cite{Johnston2020MNRAS.494.4576J} apply Markov chain Monte Carlo methods to 3840 Kepler X-ray burst models and obtain system parameter estimates for GS 1826-24. They estimate a metallicity of $Z_{\rm CNO}=0.010_{-0.004}^{+0.005}$, hydrogen fraction of $X_0=0.74_{-0.03}^{+0.02}$, mass $M>1.7M_{\odot}$, radius $R=11.3_{-1.3}^{+1.3}$, etc. So far, the NS mass and radius are unknown for burst sources, but mass and radius change the burst properties. It is worth for us to extract the information on the macroscopic properties of NS from the observation of X-ray bursts.

As X-ray burst simulations with a general relativistic stellar-evolution code solve the stellar evolution equations from the center to the surface with the EOS, neutrino emission, crust heating as well as the nuclear energy generation in accreting layers are important for the comparison to X-ray burst observations \citep{Dohi2020PTEP.2020c3E02D, Dohi2021ApJ...923...64D, 2022ApJ...937..124D}. The MESA and KEPLER code only consider the accreting layers above NS solid crust, where base heating parameter $Q_{\rm b}$ is adopted at the inner boundary to mimic the energy transfer from the NS interior. However, the value of base heating is not well constrained by observation. \cite{keek2016MNRAS.456L..11K} assumed $Q_{\rm b}=0.1 ~\rm MeV~u^{-1}$, the deep crust heating theory suggests that the generated heat may be larger, up to $Q_{\rm b}=2 \rm ~MeV~u^{-1}$ \citep{Haensel1990A&A...227..431H,Haensel2003A&A...404L..33H,Haensel2008A&A...480..459H}, although most of the heating in the deep crust is conducted into the core and carried off by neutrinos, a considerable amount of local heating will occur, which may increase $Q_{\rm b}$. 
A yet-unknown shallow heating may also increase $Q_{\rm b}$ \citep{Brown2009ApJ...698.1020B,Deibel2015ApJ...809L..31D,Lu2022RAA....22e5018L}. On the other hand, $Q_{\rm b}$ may be reduced by the competing effect of neutrino cooling \citep{Cumming2006ApJ...646..429C}, the Urca neutrino cooling process in the outer crust may also complicate the estimation of $Q_{\rm b}$ \citep{Schatz2014Natur.505...62S}. Thus it is significant for us to study the effect of $Q_{\rm b}$ on X-ray bursts, by model-observation comparison, we may give a constraint on its value. 

In addition, in Newtonian codes such as MESA or KEPLER, to accurately model bursts, it is important to account for the General Relativity (GR) effects when comparing models with observations. The MESA code adopts the Post-Newtonian correction to include the effects of GR \citep{Paxton2015ApJS..220...15P,Meisel2018ApJ...860..147M}. The KEPLER code uses Newtonian gravity and ignores the GR effects, thus GR corrections are adopted in X-ray burst simulations \citep{2011ApJ...743..189K,2018MNRAS.477.2112J,Johnston2020MNRAS.494.4576J}. In the present study, we adopt MESA code to simulate a sequence of X-ray bursts, and compare the results from a general relativistic stellar-evolution code, HERES \citep{Dohi2021ApJ...923...64D}, focusing on the burst observables.

The structure of the paper is as follows. In Section~\ref{sec:model}, we describe the Post-Newtonian hydrodynamic MESA model and the GR hydrostatic HERES model. In Section~\ref{sec:res}, we present the results of computations wherein the effects of masses, radii and base heating upon the X-ray burst properties are taken into account, and we compare the results from the MESA code and HERES code \citep{Dohi2021ApJ...923...64D} in Section~\ref{sec:com}.  Finally, we summarize our results and briefly discuss their implications.

\section{MODEL}\label{sec:model}

\begin{figure}
\centering
\includegraphics[width=\linewidth]{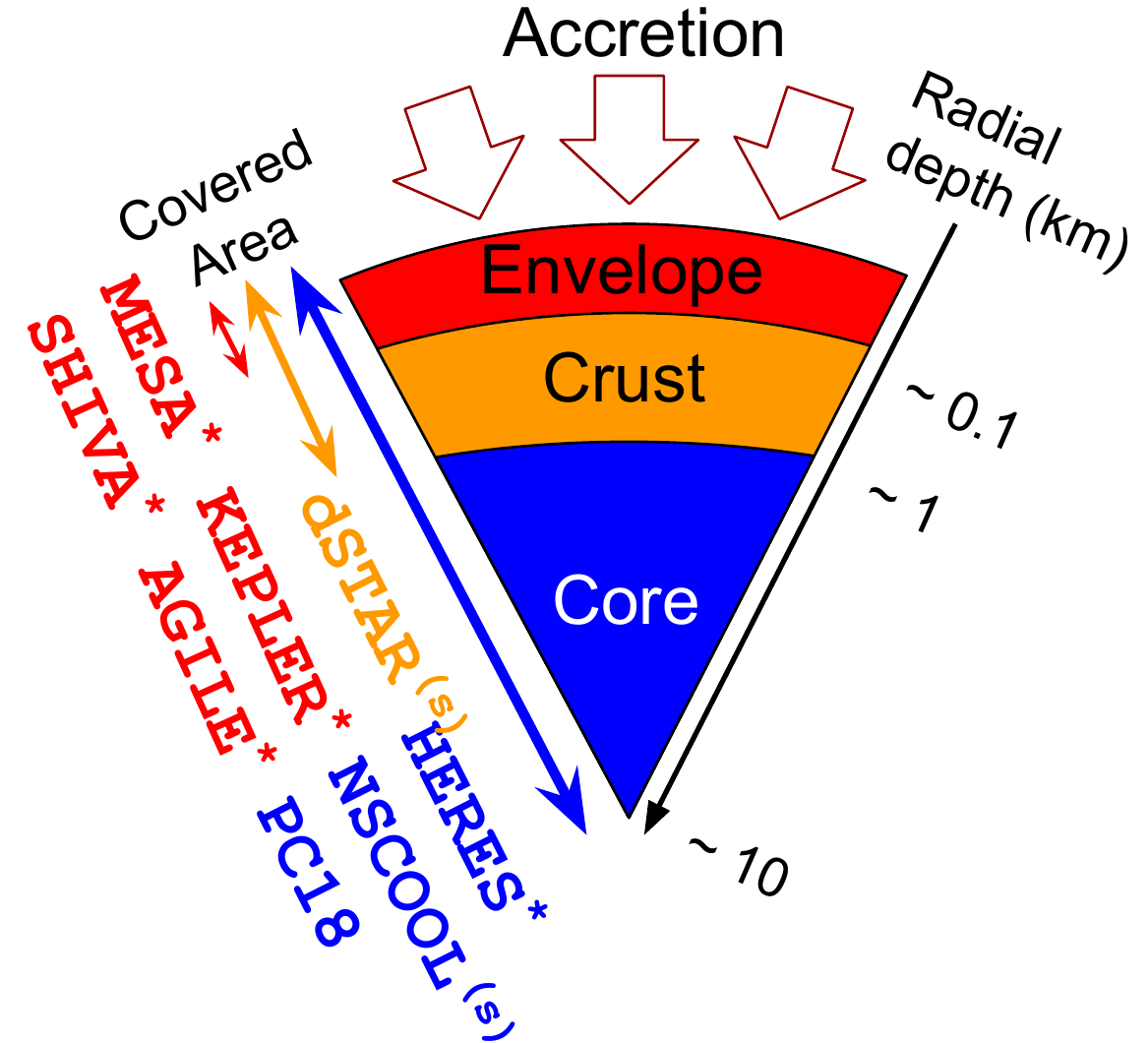}
\caption{A schematic picture of the NS structure. The computational domain of several X-ray burst codes for the thermal evolution of accreting NS is shown. The label with the asterisk (*) considers the effects of convection and nuclear reaction networks, and with (s) treats the envelope as in the steady state.}
\label{fig:nspic}
\end{figure}

There are several X-ray burst models which have different features due to the nature of burst code. In Fig.~\ref{fig:nspic}, we show the schematic of NS structure and several corresponding burst codes. In several stellar evolutionary models, the most used code is the MESA code~\citep{Paxton2011ApJS..192....3P, Paxton2013ApJS..208....4P, Paxton2015ApJS..220...15P, Paxton2018ApJS..234...34P}, which solves the (post-)Newtonian hydrodynamics within the accreted regions. The formulation of \texttt{MESA} is quite similar to some codes of \texttt{KEPLER}~\citep{Woosley2004ApJS..151...75W} and \texttt{SHIVA}~\citep{1998ApJ...494..680J,2010ApJS..189..204J}. These codes can be accessible even with the thermal evolution of relativistic compact objects by using the ``GR correction"~(e.g., \cite{2011ApJ...743..189K})\footnote{However, one of similar burst code \texttt{AGILE}~\citep{2002ApJS..141..229L,2005PhDT.......124F} is based on general relativity.}, but since the boundary condition on the crust surface is inevitably introduced as ``$Q_b$" value, it is hard for them to probe the NS physics. The approximate treatment of strong gravity in NSs as above may not be valid except for the surface, and therefore consistent treatment based on general relativistic formulation is indispensable for more exact calculation of burst light curves.

The sophisticated public code which takes into account the above is the \texttt{dSTAR} code~\citep{2015ascl.soft05034B} originally developed by \cite{2000ApJ...531..988B}. \texttt{dSTAR} simultaneously solves the TOV and energy transport equations without convection and reaction networks for X-ray bursts. It covers the regions except for the NS core and can probe the crust physics such as crustal heating, shallow heating, Urca cooling, and so on~\citep{Deibel2015ApJ...809L..31D,Deibel2016ApJ...831...13D,Deibel2017ApJ...839...95D,Meisel2017ApJ...837...73M}. For the envelope, it treats as in the steady state, which can construct a relation between surface temperature and the crust temperature at the shallowest point~\citep{2002ApJ...574..920B}. Still, \texttt{dSTAR} leaves the boundary condition on the core surface, which must be changed by the NS physics such as the EOS and $\nu$ cooling effects.

To include more possible NS physics, we have recently developed the code of \texttt{HERES}\footnote{The name derives from ``One-Dimensional Hydrostatic Evolution of RElativistic Stars". Our code originally derives from \cite{Fujimoto1984ApJ...278..813F}.}~\citep{Dohi2020PTEP.2020c3E02D}. \texttt{HERES} is essentially the same as \texttt{dSTAR} in that both follow quasi-thermal evolution of accreting NSs, but the covered regions for calculation are extended to the center of the NSs. Unlike the other codes, no artificial boundary condition such as $Q_b$ is required. Another two codes of \texttt{NSCool}~\citep{2001ApJ...548L.175C,2013PhRvL.111x1102P}\footnote{Updated code for accreting NSs from the original one~\citep{1989PhDT........45P}. The envelope is treated in steady state.} and \texttt{PC18}~\citep{2018A&A...609A..74P}\footnote{Note that effects of the magnetic field in NSs is considered, unlike other codes.} (see also \cite{2021A&A...645A.102P}) are similar to \texttt{HERES} in regard with the formulation but without convection and reaction network for X-ray bursts. Therefore, \texttt{HERES} is currently the unique code that can probe the NS physics from X-ray burst light curves. In this work, we adopt two distinct codes of \texttt{MESA} and \texttt{HERES}. Next, we briefly explain the properties of each code.


\subsection{Post-Newtonian Hydrodynamic model with large reaction network (MESA)}
\label{subsec:mesa}

We use an open-source stellar evolution code (\texttt{MESA}, version 9793)~\citep{Paxton2015ApJS..220...15P} to perform calculations on Type I X-ray bursts. The MESA equation of state (EOS) is based on the 2005 OPAL EOS tables \citep{2002ApJ...576.1064R}, besides, the SCVH tables \citep{1995ApJS...99..713S}, HELM \citep{2000ApJS..126..501T} and PC \citep{2010CoPP...50...82P} EOSs are employed for various conditions \citep{Paxton2011ApJS..192....3P}. It is worth to mention that a new Skye EOS for fully ionized is designed by \cite{2021ApJ...913...72J}, which has been tested in action in the MESA stellar evolution code by computing white dwarf cooling curves. 
OPAL opacity tables are used with the proto-solar abundances from \cite{2009ARA&A..47..481A}. 
Following the approach described in \cite{Meisel2018ApJ...860..147M}, we model a series of NS envelopes by considering inner boundary conditions for different NS masses and radii. The most pertinent details are repeated here. The luminosity at the base of the envelope is set to $L_{\rm base}=\dot{M}Q_{\rm b}$, where $Q_{\rm b}$ is the base heat, a parameter adopted by many models to simulate the heat flow from the neutron star's crust into the envelope\citep{Brown2009ApJ...698.1020B, Galloway2021ASSL..461..209G, Keek2017ApJ...842..113K}. The change of mass, radius and base luminosity by using the commands ``\textit{relax\underline{~}M\underline{~}center}'', ``\textit{relax\underline{~}R\underline{~}center}" and ``\textit{relax\underline{~}L\underline{~}center}", respectively \citep{Paxton2011ApJS..192....3P}. GR effects were accounted for using a post-Newtonian modification to the local gravity \citep{Paxton2011ApJS..192....3P,Paxton2015ApJS..220...15P}, where the MESA setting ``\textit{use\_GR\_factors = .true.}" was chosen. The envelope thickness is approximately $0.01~\rm km$, and the initial metal abundance uses the solar metal abundance $Z=0.01, 0.02$ \citep{Grevesse1998SSRv...85..161G}. We use the rp.net, which contains 304 isotopes (see \citet{Fisker2007astro.ph..3311L}), and nuclear reaction rates use the reaction rates from the REACLIB V2.2 library\citep{Cyburt2010ApJS..189..240C}. Adaptive time and spatial resolution were employed according to the MESA controls \textit{varcontrol\underline{~}target=1d-3} and \textit{mesh\underline{~}delta\underline{~}coeff=1.0} \citep{Paxton2013ApJS..208....4P}. In order to achieve convergent solutions, some models need slightly different settings. In table \ref{tab:p} from the Appendix, we provide our burst models, which describe the input parameters and some of the outputs in more detail. 

\subsection{General-Relativistic hydrostatic Evolutional model with an approximate reaction network (\texttt{HERES})}
\label{subsec:heres}

As explained above, \texttt{MESA} has two issues of the treatment of NS gravity and artificial boundary condition introduced as $Q_{\rm b}$. As the definitions of $Q_{\rm b}$ are different in previous works, e.g., $Q_{\rm b}$ is defined by \cite{keek2016MNRAS.456L..11K} to mean the amount of heat generated by crustal heating at the base of the envelope, and the typical value for $Q_{\rm b}$ of 0.1 MeV/u was adopted. $Q_{\rm b}$ is defined by \cite{Meisel2018ApJ...860..147M} to mean not only the crust heating but also the shallow heating. $Q_{\rm b}=0.1,0.5,1.0 ~\rm MeV/u$ were adopted to mimic the shallow heating of unknown origin. Hereafter, we define the net base heat as $Q_{\rm e}$, which represents the energy exchange between the interior NS and the accreting layer, its value could be changed by the unknown shallow heating or the $\nu$ cooling processes inside NSs, related to the EOS and mass~(see Table 2 in \cite{Dohi2021ApJ...923...64D}). In such formulation, it is principally impossible to treat
the heat flux coming from the interior of NSs, which drastically changes the overall temperature through electron (and radiative) thermal conductivity. Thus, we should validate the \texttt{MESA} burst models in particular for the physical effects inside NSs.

As the most realistic burst model which covers whole NS regions, we utilize some of them presented by \cite{Dohi2021ApJ...923...64D}, which follows quasi-hydrostatic evolution by using \texttt{HERES}. We take an approximate reaction network with 88 isotopes for mixed hydrogen and helium burning~(APRX3 in \cite{Dohi2020PTEP.2020c3E02D}) and the same data of reaction rates as MESA \footnote{Regarding ${}^{64}{\rm Ge}(p,\gamma){}^{65}{\rm As}$ and ${}^{65}{\rm As}(p,\gamma){}^{66}{\rm Se}$, \cite{Dohi2021ApJ...923...64D} adopted the data from \cite{2016ApJ...818...78L}. In this paper, however, we re-make HERES burst models with their reaction rates of \cite{Cyburt2016ApJ...830...55C}, which is implemented in MESA. 
}. In the energy transport equation, we implement the Schwarzschild convection. Note that convection is required for causing the mixed hydrogen/helium burning though it is somewhat artificial due to one-dimensional formulation. The initial models for our X-ray burst calculation are set to be the steady-state models~\citep{2021PhRvD.103f3009L} with gravitational compressional heating~(see \cite{2018IJMPE..2750067M} for details).


Let us explain model parameters in \texttt{HERES}. The accretion rate and compositions of accreted matter are the same as Sec~\ref{subsec:mesa}. We utilize the nuclear EOS of Togashi, which has been based on the variational approach with the use of the bare nuclear potentials for two-body interaction and phenomenological three-body
interaction~\citep{2017NuPhA.961...78T}. For the heating source, standard crustal heating rates of \cite{Haensel1990A&A...227..431H} are implemented. For the cooling source, we consider the slow $\nu$ cooling processes mainly composed of the modified Urca process and bremsstrahlung. The occurrence of fast $\nu$ cooling processes such as the nucleon direct Urca process, i.e., neutrino emissions induced by (inverse) $\beta$ decay, could affect burst light curves~\citep{2022ApJ...937..124D}, but for any mass, it is prohibited with the Togashi EOS due to the quite low symmetry energy (the slope parameter $L$ is 30 MeV)~\citep{2019PTEP.2019k3E01D}.




\section{The impact of neutron star mass, radius and base heating on type I X-ray burst} \label{sec:res}
We build a series of scenarios (models 1-12 in Table \ref{tab:p}) with variation in mass (models 1-4), radius (models 5-8) and base heating (models 9-12), then type I X-ray bursts on the surface layer of accreting NSs are simulated by using MESA with the above inputs.

The light curves of X-ray bursts are usually characterised by several parameters, e.g., the recurrence time $\Delta t$, which represents the time from one burst to the next. The burst duration $\tau$, which is defined to be the time after the peak at half value of $L_{\rm peak}$. The rise time $t_{\rm rise}$ is defined from transience to peak point. The e-folding time $\tau_e$\ is defined after peak point. The peak luminosity $L_{\rm peak}$ is taken from the light curve maximum. The burst energy $E_b$ is obtained by integrating over the light curve
\begin{equation} \label{eq:eb}
E_b=\int L_b d t
\end{equation}
The burst strength $\alpha$ is defined by the ratio of the accretion energy to the burst energy
\begin{equation}\label{eq:alpha}
\alpha=\frac{z_{g}}{1+z_{g}} \dot{M} c^{2} \frac{\Delta t}{E_{b}}
\end{equation}
where $z_{g}$ is the gravitational redshift.

In order to compare with observations, we stack a sequence of bursts from each model and obtain the average light curve, burst parameters and $1\sigma$ error for them. Since the wait time for the next burst is usually shortened as the ash from the previous burst is mixed with the new fuel, i.e., \textit{compositional inertia}~\citep{1980ApJ...241..358T,Woosley2004ApJS..151...75W}, we remove the data of the first four bursts and start processing from the fifth burst. The convergence of MESA light curves is almost archived at $\sim$5 bursts, which is fewer than $\sim$10 bursts in KEPLER without nuclear preheating~(see Figure A1 in \cite{Johnston2020MNRAS.494.4576J})\footnote{\texttt{HERES} light curves are converged around 10--30 burst times, which are more than those in MESA and KEPLER. This is because the HERES adopts the aniso-thermal structure as the initial model~\citep{2018IJMPE..2750067M}, which spends the convergence time due to the existence of thermal flux.}.

\subsection{Variations in NS mass, radius and base heating, and X-ray burst parameters} \label{sec:exp}

\begin{figure}[b]
\centering
	\includegraphics[width=\columnwidth]{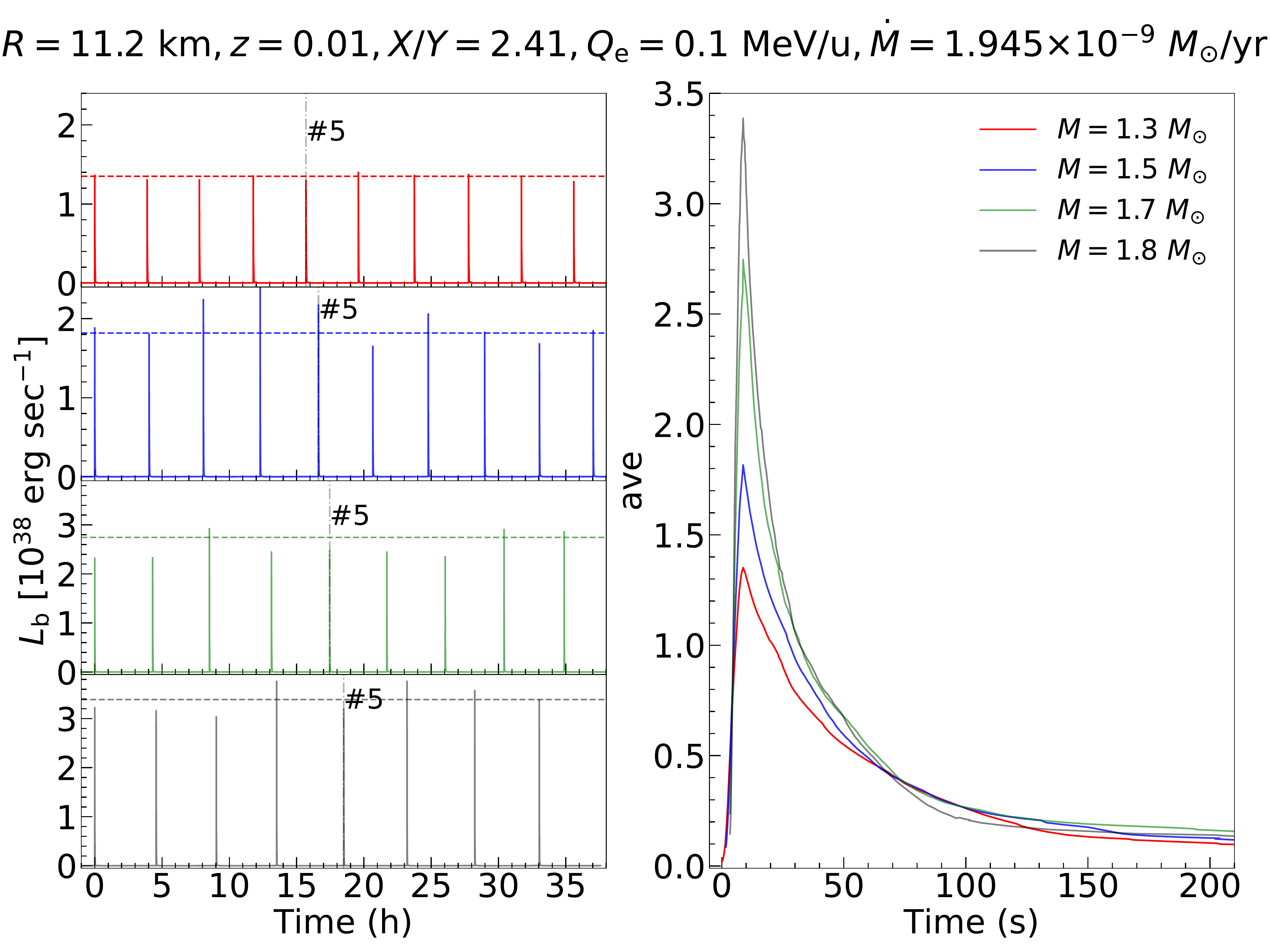}
    \caption{Light curves with different mass models. Left: The luminosity of the burst sequence during 0–40 h with different NS mass models, i.e., 1.3$ M_{\odot}$ (black), 1.5$ M_{\odot}$ (green), 1.7$ M_{\odot}$ (blue), and 1.8$ M_{\odot}$ (red). The horizontal dashed line in each panel represents the average peak luminosity. Right: Average light curves.}
    \label{fig:LM}
\end{figure}

\begin{figure}[t]
	\centering
		\includegraphics[width=\columnwidth]{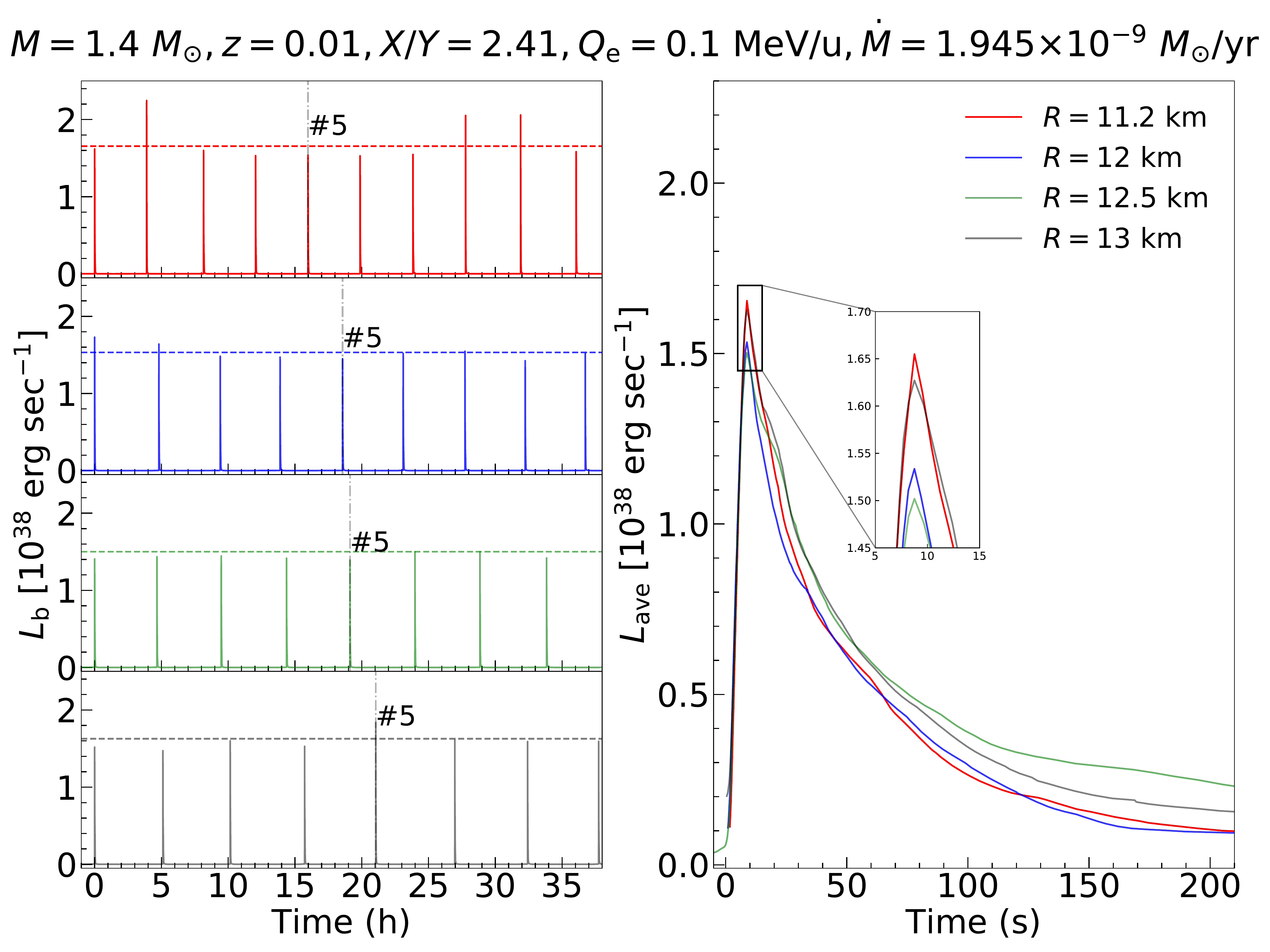}
    	\caption{Light curves with different radius models. Left: The luminosity of the burst sequence during 0–40 h with different NS radius models. i.e., 11.2 km (black), 12 km (yellow), 12.5 km (blue), and 13 km (red). The horizontal dashed line in each panel represents the mean peak luminosity. Right: Average light curves.}
    	\label{fig:LR}
	\end{figure}

\begin{figure}[t]
\centering
	\includegraphics[width=\columnwidth]{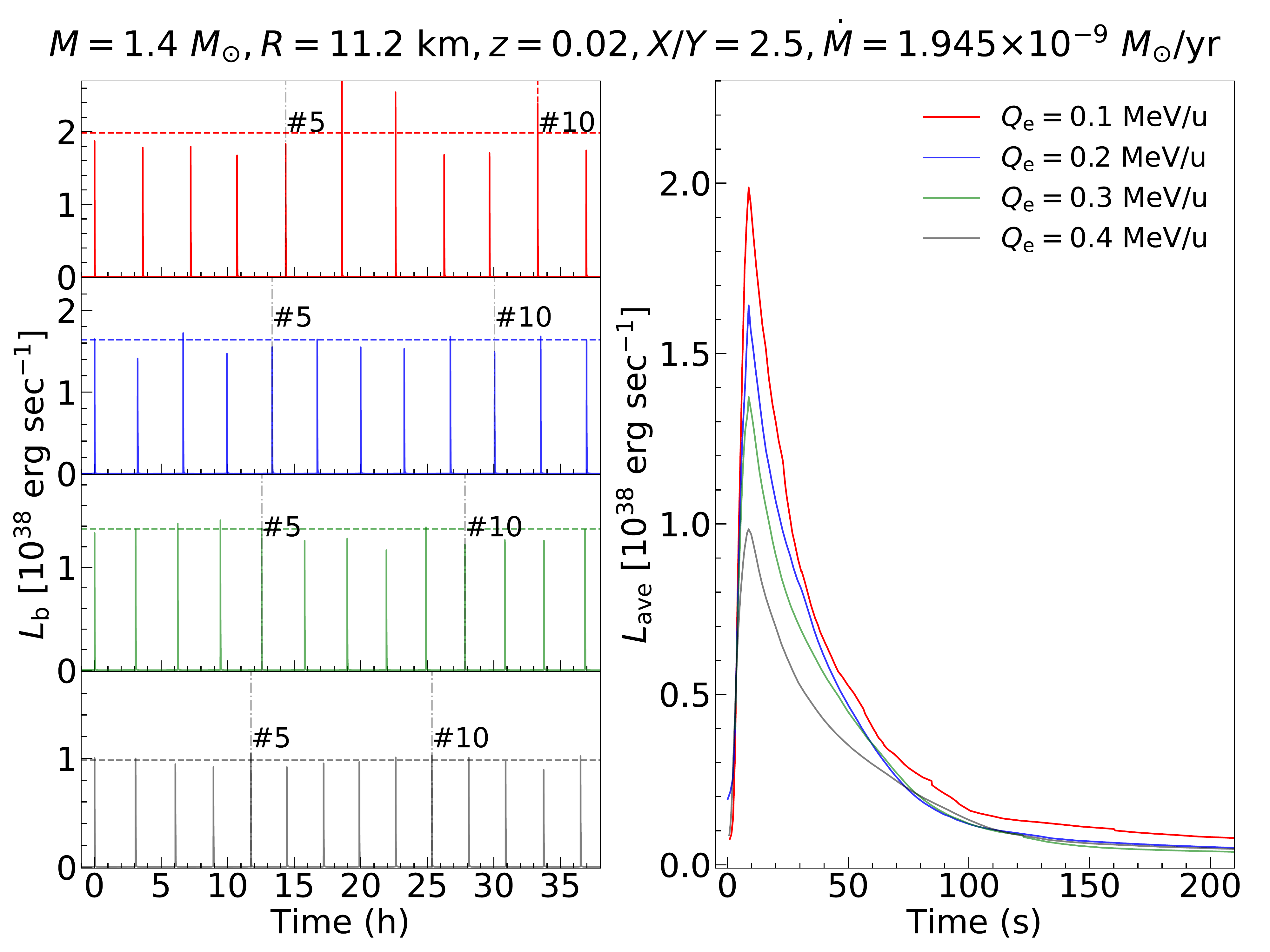}
    \caption{Light curves with different $Q_{\mathrm{e}}$ . Left: The luminosity of the burst sequence during 0–40 h with different $Q_{\mathrm{e}}$. i.e., 0.1 MeV/u (black), 0.2 MeV/u (yellow), 0.3 MeV/u (blue), and 0.4 MeV/u (red). The horizontal dashed line in each panel represents the mean peak luminosity. Right: Average light curves. }
    \label{fig:LQ}
\end{figure}

X-ray burst with various values of NS mass, radius and base heating are calculated. In the left panel of Figure \ref{fig:LM}, we show the luminosity of the burst sequence with different NS mass models. We calculate the averaged light curves by aligning bursts in each sequence by their peak luminosities, the results are shown in the right panel. Similarly, the luminosity of the burst sequence with different NS radius models are shown in the left panel of Figure \ref{fig:LR} and the averaged light curves are shown in the right panel.  We find that with the increase of mass, $L_{\mathrm{peak}}$  increases, $\Delta t$  increases, decay time decreases. However, as radius is increased, $\Delta t$ also increases, $L_{\mathrm{peak}}$ remains constant, decay time increases. The results for different base heating cases are shown in  Figure \ref{fig:LQ}, as $Q_{\rm e}$ is increased, $L_{\rm peak}$ decreases, $\Delta t$ decreases and decay time decreases. In the following, we calculated burst parameters such as $\Delta t$, $L_{\rm peak}$, $\alpha$, $E_{\rm burst}$, $t_{\rm rise}$, $\tau$ and $\tau_{\rm e}$, one can find the values in detail in Table \ref{tab:p}. Meanwhile, the ignition pressure $P_{\rm ign}$ for each model is obtained to understand the variation of parameters.


\begin{figure}[b]
	\centering
		\includegraphics[width=\columnwidth]{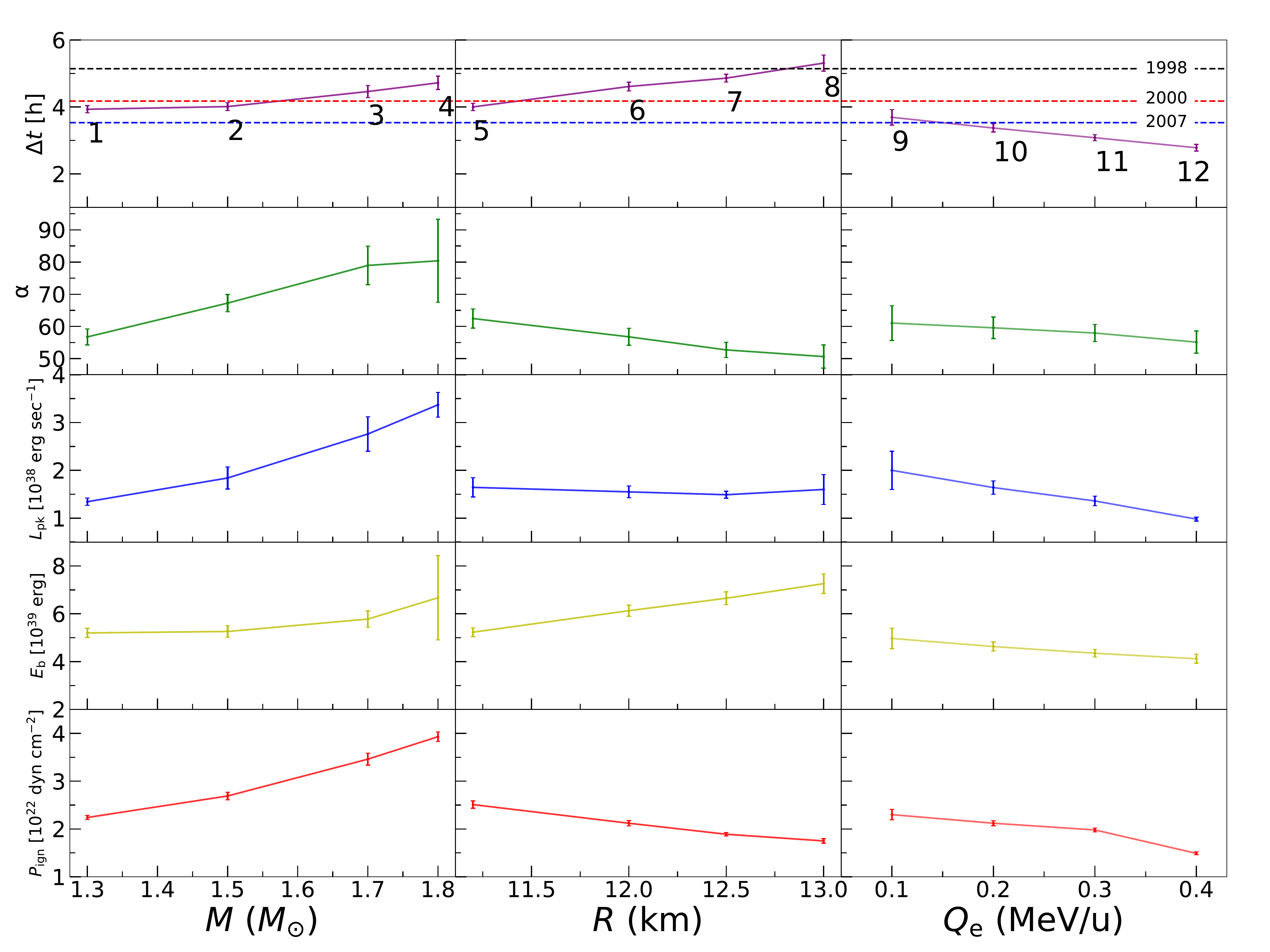}
    	\caption{The values of $\Delta t$, $\alpha$, $L_{\mathrm{peak}}$, $E_{\rm b}$ and $P_{\mathrm{ign}}$ with different $M$(left), $R$(middle) and $Q_{\rm e}$(right). The dashed lines in the uppermost panel indicate the recurrence time from observations. The numbers 1-12 in the uppermost panel indicate the model number which is shown in table \ref{tab:p} from appendix.}
    	\label{fig:PMR}
	\end{figure}

The parameters change with the variation in mass, radius and base heating are shown in Figure \ref{fig:PMR}. For the models 1-4 in Table 1 (left panel in Figure \ref{fig:PMR}), as $M$ increases, $\Delta t$, $\alpha$, $L_{\mathrm{peak}}$, $E_{\mathrm{b}}$ and $P_{\mathrm{ign}}$ increase. For a fixed NS radius, $M$ increases, the surface gravitational acceleration ($g_{\rm s}$) will become larger, resulting in an increase in ignition pressure($P_{\mathrm{ign}}$). One can also find the ignition pressure from the bottom panel in Figure \ref{fig:PMR}, which is increased with mass increases. According to one-zone model, the column density $\sigma$ is expressed in two ways(\cite{Bildsten1998ASIC..515..419B,2022ApJ...937..124D}):
\begin{equation}
    \sigma=\frac{\dot{M} \times \Delta t}{4\pi R^2}=P_{\rm {ign}}/g_{\rm s}
\end{equation}
where $g_{\rm s}=\frac{GM}{R^2}(1-\frac{2GM}{Rc^2})^{-1/2}$, one can see the surface gravity acceleration $g_{\rm s}$ on the mass-radius plane in detail from Figure 3 of \cite{Dohi2021ApJ...923...64D}.
As a result, for the fixed accretion rate and NS radius, with increasing mass, the recurrence time is proportional to $P_{\rm {ign}}/g_{\rm s}$. The increase of ignition pressure overtakes the increase of surface gravity acceleration, which leads to the increase of $\Delta t$. The peak luminosity can be scaled as the Eddington limit \citep{Lewin1993SSRv...62..223L}, 
\begin{equation}
\label{fig:col}
 L_{\rm peak}\sim L_{\rm Edd}=4\pi cGM/\kappa \propto M
\end{equation}
which is proportional to $M$, but independent of $R$. where $\kappa$ is the electron scattering opacity. Therefore, the peak luminosity is increased with $M$ increases.

Assumed all accreted matter is processed in flashes, the burst strength is the ratio of the average luminosity emitted in the persistent X-ray emission ($L_{\rm p}$) to that emitted in X-ray bursts ($L_{\rm b}$)\citep{Lewin1993SSRv...62..223L}:
\begin{equation} \label{eq:alp2}
\alpha=\frac{L_{\rm p}}{L_{\rm b}}=\frac{\varepsilon_{\rm G}}{\varepsilon_{\rm N}}\sim(25-100)\frac{M/M_{\odot}}{R/10~{\rm km}}
\end{equation}
where $\varepsilon_{\rm G}=GM/R$ is the gravitational energy release per gram, $\varepsilon_{\rm N}$ is the nuclear energy. According to equation (\ref{eq:alp2}), for a fixed NS radius, $\alpha$ increases as mass increases. Our results from MESA simulation are almost consistent with the above simple one-zone model assumption.

In the middle panel of Figure \ref{fig:PMR}, it shows that $P_{\mathrm{ign}}$ and $\alpha$ of the bursts are inversely proportional to the radius, and $\Delta t$ and $E_{\mathrm{b}}$ are proportional to the radius. While the peak luminosity $L_{\rm peak}$ remains constant. As the radius increases, the gravitational acceleration on the surface of the neutron star becomes smaller, the ignition pressure decrease $\Delta t$ is longer due to the increased NS surface area. Burst energy is larger due to the longer e-folding time, to the reduced gravitational redshift from the NS surface. Burst strength $\alpha$ is reduced due to the lower surface gravitational potential, which also can be easily understood from equation \ref{eq:alp2}. According to equation \ref{fig:col}, as the peak luminosity does not dependent on radius, the peak luminosity almost constant with radius increases.   

 The results for the parameter variation with base heating are shown in the right panel of Figure \ref{fig:PMR}. With the increase of $Q_{\mathrm{e}}$, the peak luminosity of the burst decreases continuously, and the interval between bursts becomes smaller.
 This is because the first hot CNO cycle, i.e., ${}^{\rm 12}{\rm C}(p,\gamma){}^{\rm 13}{\rm N}(p,\gamma){}^{\rm 14}{\rm O}(\beta^+){}^{\rm 14}{\rm N}(p,\gamma){}^{\rm 15}{\rm O}(\beta^+){}^{\rm 15}{\rm N}(p,\alpha){}^{\rm 12}{\rm C}$, lasts longer with smaller $Q_{\rm e}$; The timescale of hot CNO cycle is almost determined by abundances of ${}^{14}{\rm O}$ and ${}^{15}{\rm O}$, which could trigger new $(\alpha,p)$ reaction paths, indirectly leading to proton-rich nucleosynthesis
\footnote{At low temperature of $T\lesssim4\times10^8~{\rm K}$, $\beta$ decays are dominant, but at high temperature, the second hot CNO cycle, ${}^{\rm 14}{\rm O}(\alpha,p){}^{\rm 17}{\rm F}(p,\gamma){}^{\rm 18}{\rm Ne}(\beta^+){}^{\rm 18}{\rm F}(p,\alpha){}^{\rm 15}{\rm O}$, occurs instead of ${}^{14}{\rm O}(\beta^+)$. The resultant breakout reactions to $\alpha p$ and $rp$ processes are therefore ${}^{15}{\rm O}(\alpha,\gamma){}^{19}{\rm Ne}$ and ${}^{18}{\rm Ne}(\alpha,p){}^{21}{\rm Na}$.}. If $Q_{\mathrm{e}}$ is smaller (in the range of $0<Q_{\rm e}<$0.5 MeV/u), i.e., interior NS is colder, excessive protons turn into helium, which burns to ${}^{\rm 12}{\rm C}$ by the $3\alpha$ reaction at faster rates because it takes more time to accumulate the seeds, ${}^{14}{\rm O}$ and ${}^{15}{\rm O}$, leading to higher $\Delta t$. Then, e-folding time tends to be shorter because protons being critical fuel of the $rp$ process are more exhausted.  As a result, the larger energy is produced during the hot CNO cycle due to its longer duration if $Q_{\rm e}$ is smaller, leading to a higher peak luminosity. We note that $Q_{\rm e}$ dependence of the hot CNO cycle timescale is, in a sense, similar to the ${}^{15}{\rm O}(\alpha,\gamma){}^{19}{\rm Ne}$ rate dependence of that, which has been studied by \cite{2006ApJ...650..332F,2007ApJ...665..637F}.


\subsection {Model-observation comparisons}	

\begin{figure}[b]
\centering
	\includegraphics[width=\columnwidth]{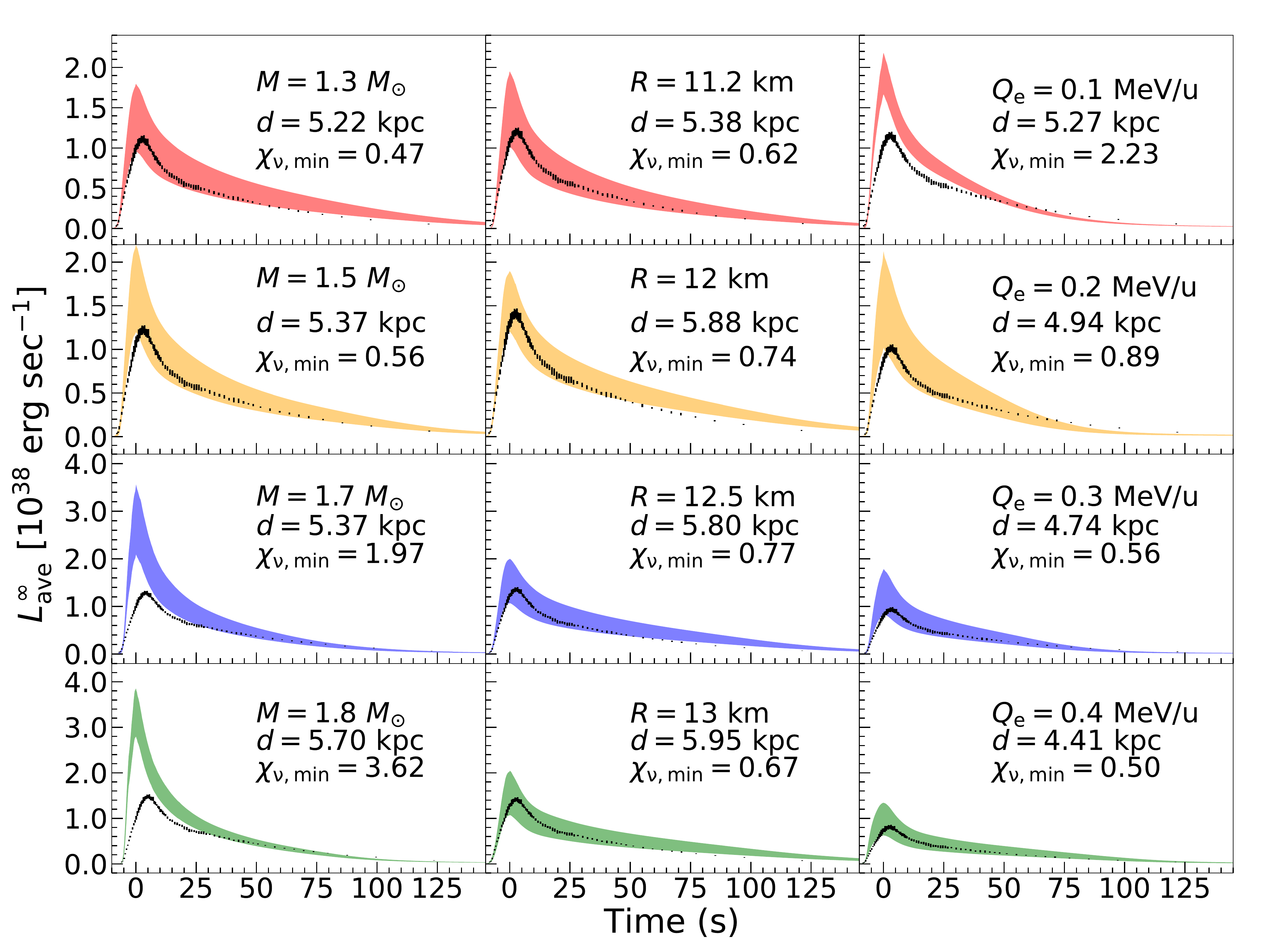}
	\caption{Comparison of the calculated averaged burst light curves with $1\sigma$ error regions ($M=1.3,1.5,1.7,1.8 M_{\odot}$ and $R=11.2,12,12.5,13 ~\rm km$ and $Q_{\rm e}=0.1,0.2,0.3,0.4 ~\rm MeV/u$) with observed ones of GS1826-24 in 2007.}
    \label{fig:OMRQ}
\end{figure}
The light curves with variations in NS mass, radius and base heating are compared with observation in Figure \ref{fig:OMRQ}, where the observed light curve of GS 1826-24 in 2007 is adopted.  We include the burst anisotropy $\xi_b$ in the distance, $d\xi_{b}^{1/2}$ is calculated from $F_{\rm peak}=L_{\rm peak}/4\pi d^2\xi_{b}$. With use of the $\chi^2$ method in \cite{Dohi2020PTEP.2020c3E02D}, we can get the best fit $d\xi_{\rm b}^{1/2}$ for each model-observation comparison. From the left panel of Figure \ref{fig:OMRQ}, we can see that the peak luminosity increases as mass increases, the peak luminosity is too high to fit the observation for $M\geq1.7M_{\odot}$. In the middle panel of Figure \ref{fig:OMRQ}, the peak luminosity almost constant as radius increases, the light curve can be well fitted with radius in the range $\sim$11.2-13 km.  In the right panel of figure \ref{fig:OMRQ}, the peak luminosity decreases as base heating increases,  the light curve can be well fitted with the variation of base heating in the range $Q_{\rm e}=0.1-0.4~\rm MeV/u$. Besides, we also compare the recurrence time with observation in the uppermost panel in figure \ref{fig:PMR}. The burst models of 1-10 are consistent with observed values. The recurrence time is too short to interpret observations for burst models 11-12 with $Q_{\rm e}=0.3 , 0.4 ~\rm MeV/u$. However, as the source distance is uncertain, which is crucial to determine the shape of the light curve. The input parameters such as metallicity, accretion rate also affect the burst light curve. It is better for us to use the MCMC method (e.g. \cite{Johnston2020MNRAS.494.4576J}) to determine the system parameters. In our calculation, models 9 and 10 are consistent with the observation of GS 1826-24 in 2007 (whether the light curve or the recurrence time).

\section{Code comparison} \label{sec:com}
  
\begin{figure}[htbp]
\centering
	\includegraphics[width=\columnwidth]{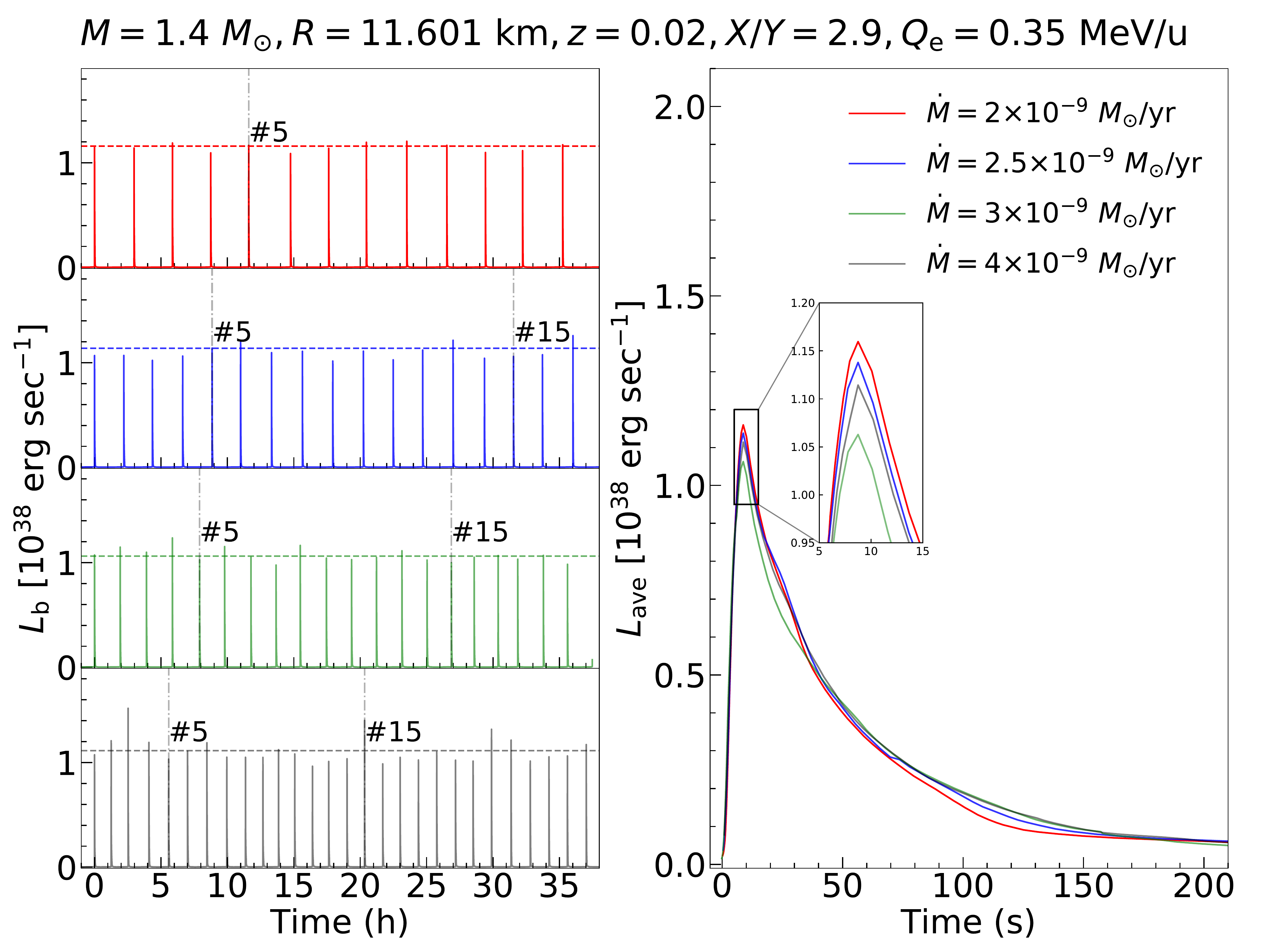}
   \caption{Light curves under different $\dot{M}$ with use of HERES code. Left: The luminosity of the burst sequence during 0–40 h for several $\dot{M}$. The horizontal dashed line in each panel represents the mean peak luminosity. Right: Average light curves.}
    \label{fig:LHer}
\end{figure}

In order to validate the models for X-ray burst calculation such as MESA which solves the Newtonian hydrodynamics with the accreted layers, we adopt the realistic code \texttt{HERES} which solves the whole NS as a comparison. In table \ref{tab:Heres} from the appendix, we show our calculation models with HERES code. By using adopted mass, radius and accretion rate under $X/Y=2.9$ and $Z_{\rm CNO}=0.02$, we obtain several burst parameters, such as burst strength $\alpha$, burst duration $\tau$, recurrence time $\Delta t$, total burst energy $E_{\rm burst}$, peak luminosity $L_{\rm peak}$, rise time $t_{\rm rise}$. The $1\sigma$ errors are also presented for each output parameters. The base heating inferred from the $1.4M_{\odot}$ NS with Togashi EoS is $Q_{\rm e}=$0.35 MeV/u (\cite{Dohi2021ApJ...923...64D}). The light curves calculated with the HERES code are shown in Figure \ref{fig:LHer}. It shows that the recurrence time decreases as accretion rate increases, the peak luminosity almost constant. Meanwhile, we adopt the same mass, radius, metallicity, $X/Y$, base heating, accretion rate for MESA X-ray burst calculations. The input parameters and some of the output parameters are shown in table \ref{tab:p} from model 13 to model 16 in appendix.

\begin{figure}[t]
\centering
	\includegraphics[width=\columnwidth]{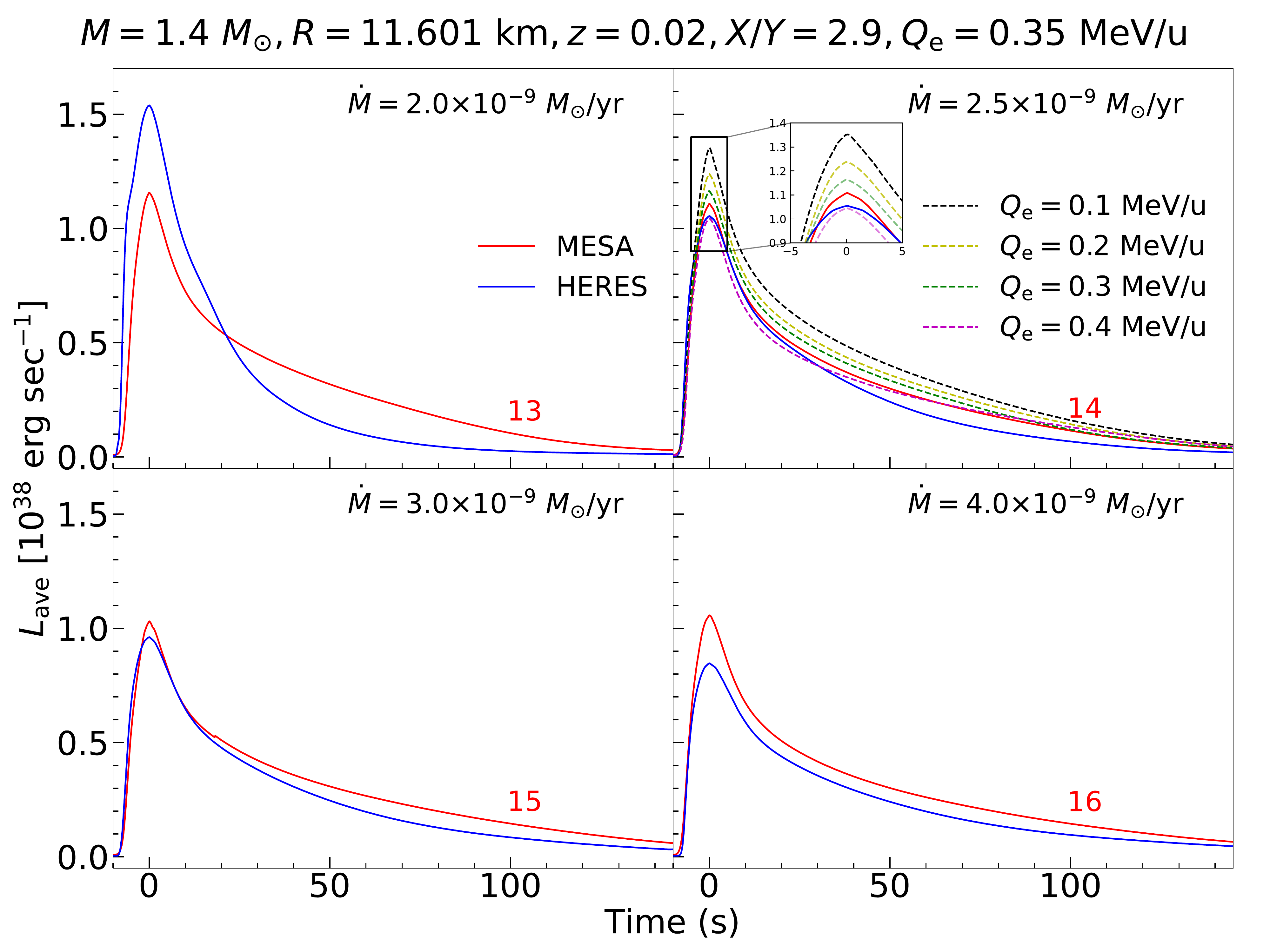}
	\caption{The mean burst light curve calculated from MESA code (red solid line and the model numbers ``13"-``15'' are marked in each panel) versus the mean light curve calculated from HERES code (blue solid line), where $Q_{\rm e}=0.35~\rm MeV/u$. In the upper right panel, the dashed lines in different colors (black, yellow, green, purple) are same as the red solid line but with different base heating (0.1, 0.2, 0.3, 0.4 MeV/u).}
    \label{fig:vs}
\end{figure}

\begin{figure}[t]
\centering
	\includegraphics[width=\columnwidth]{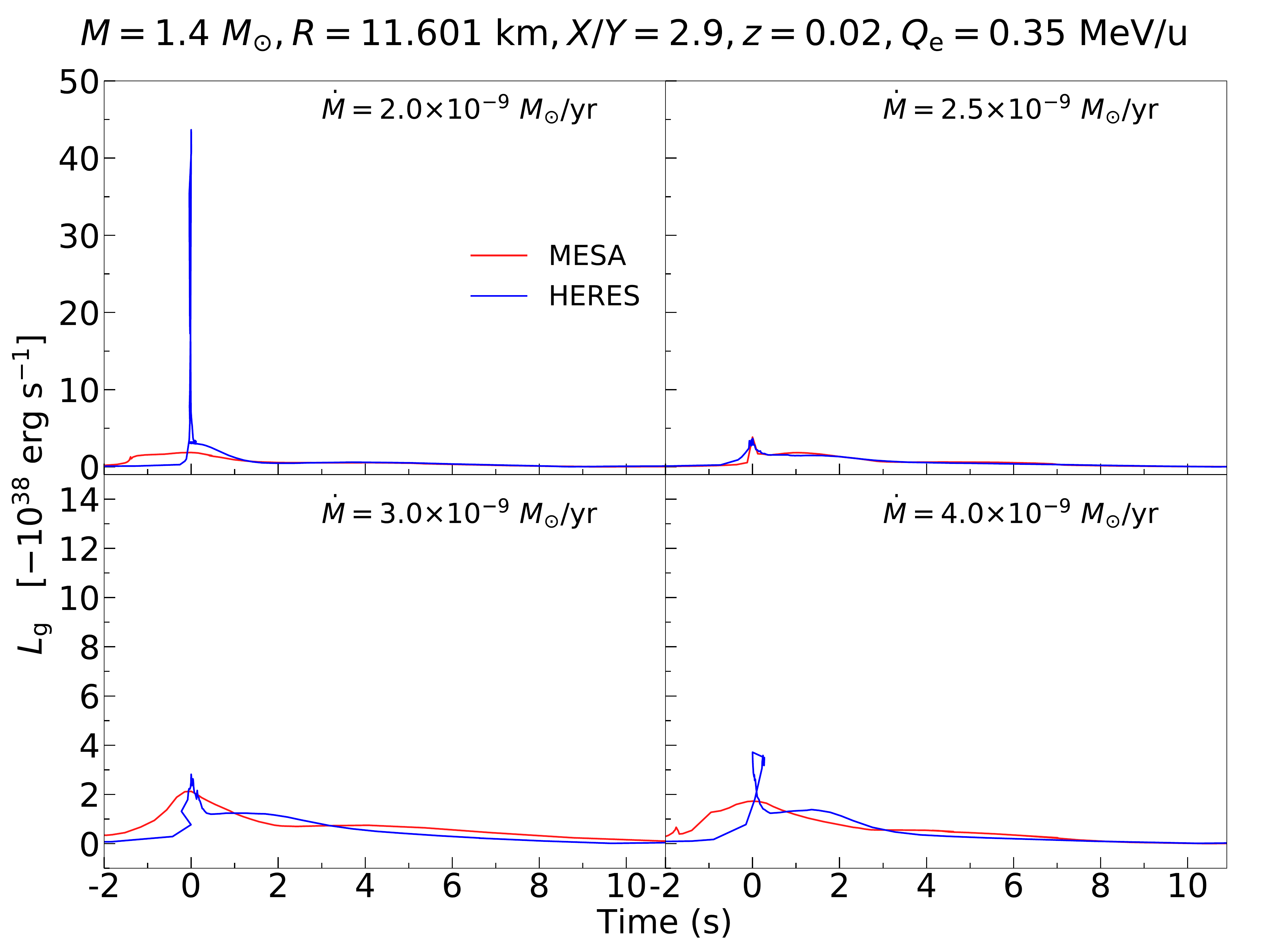}
    \caption{Comparison of the compressional heating luminosity $L_{\rm g}$ for the MESA code (red) and HERES code (blue). } 
    \label{fig:lgvs}
\end{figure}

Figure \ref{fig:vs} shows the comparison of the mean light curves between MESA and HERES calculation. The difference between two light curves is very small with accretion rate $\dot{M}=2.5\times 10^{-9}~ M_{\odot}~\rm yr^{-1}$ and $\dot{M}=3.0\times 10^{-9}~ M_{\odot}~\rm yr^{-1}$. There are big differences for peak luminosity and luminosity at the tail parts of the light curve between two codes under accretion rate $\dot{M}=2.0\times 10^{-9}~ M_{\odot}~\rm yr^{-1}$ and $\dot{M}=4.0\times 10^{-9}~ M_{\odot}~\rm yr^{-1}$.  The main difference is due to the higher hydrostatic force, i.e., higher compressional heating in \texttt{HERES} models~ \citep{2018IJMPE..2750067M}, which leads to higher peak luminosity than \texttt{MESA}.
Note that the contribution of compressional heating to total luminosity ($\sim10^{38}~{\rm erg/s}$) is around 10\%. We show in Figure \ref{fig:lgvs} that the differences of compressional heating luminosity $L_{\rm g}$ between MESA and HERES codes with 4 different accretion rates. At low accretion rate $\dot{M}=2.0\times 10^{-9}~ M_{\odot}~\rm yr^{-1}$, the peak luminosity of $L_{\rm g}$ obtained from HERES code is much higher than that in MESA code. While for the rest three accretion rates, compressional heating luminosities are almost the same between two codes.
The difference due to reaction networks, i.e., nuclear burning energy rates and compositions, appears
in the tail parts, where the luminosity is higher in \texttt{MESA} models regardless of $\dot{M}$.

Next, we also calculate models 17-20 with different base heating based on model 14 in the right upper panel in Figure \ref{fig:vs}, it shows that the lower the base heating, the higher the peak luminosity and the luminosity at the tail parts, which leads to a big deviation from the case with $Q_{\rm e}=0.35~\rm MeV/u$.

\begin{figure}[t]
\centering
	\includegraphics[width=\columnwidth]{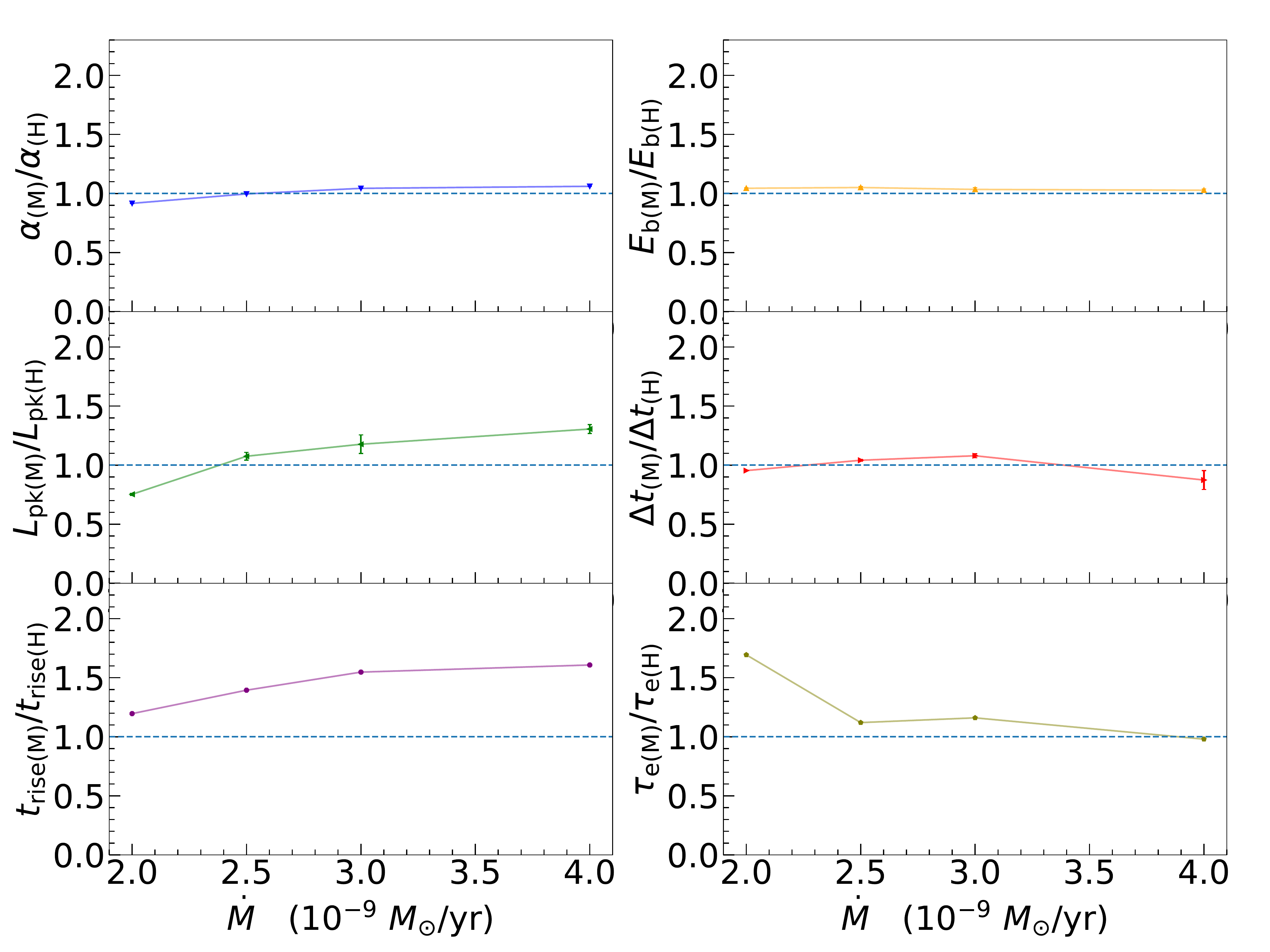}
    \caption{Comparison of the burst parameters ($\alpha$, $E_{\rm b}$, $\Delta t$, $L_{\rm peak}$, $t_{\rm rise}$, $\tau_{\rm e}$) for the MESA and HERES codes for a range of accretion rate. ``M'' and ``H'' are marked in the subscript of each parameters to indicate the results are obtained from MESA and HERES, respectively.  }
    \label{fig:pvs}
\end{figure}

Finally, we compared the predicted burst parameters ($\alpha$, $E_{\rm b}$, $\Delta t$, $L_{\rm peak}$, $t_{\rm rise}$, $\tau_{\rm e}$) of the two codes for a range of accretion rates in Figure \ref{fig:pvs}. In both codes, the burst strength $\alpha$, total burst energy $E_{\rm b}$ and recurrence time $\Delta t$ are highly consistent. The differences of the peak luminosity and the tail parts of the light curve between two codes are obvious (e.g., the maximum relative errors of $t_{\rm rise}$ and $L_{\rm peak}$ are about $\sim 50\%$). Thus, for the first time, the consistency of the two codes are identified by our comparison. The differences for the peak luminosity at low accretion rate caused by the high compressional heating luminosity as shown in Figure \ref{fig:lgvs}, while the high luminosity at the tail parts regardless of accretion rate possibly be caused by the nuclear reaction energy and compositions. The comparison of the nuclear reaction network adopted in MESA (rp.net) and HERES (APRX3) are shown in table \ref{tab:net}. It is worth noting that our input values and the values of the burst parameters obtained from MESA code which adopts a post-Newtonian modification for GR effects are unified to the local frame, in order to compare with the observations which was detected by a distant observer, the red shift of the parameters should be considered. In Appendix\,\ref{sec:red}, we show the detailed formulas to transfer the local frame quantities to the frame of a distant observer.


\section{Conclusions} \label{sec:con}

In this work, we present a set of simulations of X-ray burst with variation in NS mass, radius and base heating by using the open source code MESA. The light curves and burst parameters are obtained for each model. We find that the recurrence time, burst strength, peak luminosity and total burst energy are increased as mass increases. As radius increases, the recurrence time and total burst energy increase, the burst strength decreases, while the peak luminosity remains constant. The recurrence time, burst strength, peak luminosity and total burst energy decrease as base heating increases. The above phenomenon can be well explained with use of the simple one-zone model. One can see section \ref{sec:exp} for the detailed explanation.

The codes, such as KEPLER and MESA, solve the Newtonian hydrodynamics only within the accreted regions to simulate X-ray burst. As a result, it is hard to probe the NS physics. HERES solves the TOV and energy-transport equations, hence it can include all possible physics. To assess the validity of the boundary condition on the crust and the GR correction for the Newtonian hydrodynamics calculations, we made a comparison between multi-zone burst models from MESA and HERES code for the first time. The results show that the average light curves are highly consistent under accretion rate  $\dot{M}=2.5\times 10^{-9}~ M_{\odot}~\rm yr^{-1}$ and $\dot{M}=3.0\times 10^{-9}~ M_{\odot}~\rm yr^{-1}$. While under accretion rate $\dot{M}=2.0\times 10^{-9}~ M_{\odot}~\rm yr^{-1}$ and  $\dot{M}=4.0\times 10^{-9}~ M_{\odot}~\rm yr^{-1}$, the peak luminosity and cooling tail are obviously different between two codes. However, the burst strength, total burst energy and recurrence time are consistent between two codes regardless of accretion rate. It is worth noting that the light curves are inconsistent when we choose other values of $Q_{\rm e}$.

We demonstrate that the NS mass, radius and base heating have a non-negligible effect on the X-ray burst simulation. The validity of the boundary condition and GR correction for MESA code are verified by the code comparison between MESA and HERES. The variation trend of the output parameters with different NS mass, radius, base heating and accretion rate can help us to understand the properties of NS via X-ray burst observations.

The difference of X-ray burst codes appears in not only light curves but also $rp$-process nucleosynthesis. In fact, \cite{2008ApJS..178..110P} showed the difference in final products among three burst models with post-process calculation. A similar comparison with the use of MESA and HERES may also give information on some model parameters and will be present in near future.


\section*{Acknowledgement}
We appreciate the referee for valuable comments that improved this manuscript. We thank S. Nagataki and M. Hashimoto for their encouragement. This work received the generous support of the National Natural Science Foundation of China Nos. 12263006, U2031204, 12163005, the Natural Science Foundation of Xinjiang Province under Grant Nos. 2020D01C063, 2021D01C075 and the Science Research Grants from the China Manned Space Project with No. CMS-CSST-2021-A10. A.D. is supported by JSPS Research Fellowship for Young Scientists (22J10448). N.N. is supported by JSPS KAKENHI (19H00693, 20H05648, 21H01087) and the RIKEN Intencive Research Project.


\bibliography{sample631}{}
\bibliographystyle{aasjournal}

\appendix \label{sec:app}

\begin{table}[h] 
\centering
\tablenum{1}
\caption{Physical quantities of burst models for MESA code. Errors of output parameters indicate the $1\sigma$ standard deviation.}
\begin{tabular}[c]{cccccccccc}
    \toprule
    Model & $M$ & $R$ & $\rm Z_{CNO}$ & $\rm Q_{\rm e}$ & $\rm \dot{M}$ & $\rm \alpha$ & $\rm E_{burst}$ & $\rm L_{peak}$ & $\rm \Delta t$  \\ 
    $\rm Number$ & $ M_{\odot}$ & $\rm km$ & $ $ & $\rm MeV/u$ & $ 10^{-9}~M_{\odot}/\rm yr$ & $\rm MeV/u$ & $10^{39}~\rm erg$ & $ 10^{38}~\rm erg/s$ & $\rm h$ \\
    \hline
    1 & 1.3 & 11.2 & 0.01 & 0.1 & 1.945 & 56.70$\pm2.42$ & 5.21$\pm0.19$ & 1.34$\pm0.08$ & 3.93$\pm0.11$ \\
    2 & 1.5 & 11.2 & 0.01 & 0.1 & 1.945 & 67.24$\pm2.49$ & 5.27$\pm0.25$ & 1.81$\pm0.22$ & 4.00$\pm0.12$ \\ 
    3 & 1.7 & 11.2 & 0.01 & 0.1 & 1.945 & 79.31$\pm6.39$ & 5.79$\pm0.35$ & 2.73$\pm0.35$ & 4.48$\pm0.18$ \\ 
    4 & 1.8 & 11.2 & 0.01 & 0.1 & 1.945 & 85.30$\pm1.85$ & 6.06$\pm0.30$ & 3.37$\pm0.31$ & 4.73$\pm0.24$ \\ 
    5 & 1.4 & 11.2 & 0.01 & 0.1 & 1.945 & 62.54$\pm2.99$ & 5.23$\pm0.18$ & 1.65$\pm0.21$ & 4.00$\pm0.11$ \\ 
    6 & 1.4 & 12.0 & 0.01 & 0.1 & 1.945 & 56.30$\pm2.91$ & 6.19$\pm0.23$ & 1.54$\pm0.09$ & 4.61$\pm0.47$ \\ 
    7 & 1.4 & 12.5 & 0.01 & 0.1 & 1.945 & 52.76$\pm2.47$ & 6.67$\pm0.26$ & 1.50$\pm0.08$ & 4.87$\pm0.12$ \\ 
    8 & 1.4 & 13.0 & 0.01 & 0.1 & 1.945 & 50.77$\pm3.70$ & 7.30$\pm0.41$ & 1.62$\pm0.33$ & 5.34$\pm0.23$ \\
    9 & 1.4 & 11.2 & 0.02 & 0.1 & 1.945 & 61.05$\pm5.40$ & 4.97$\pm0.43$ & 2.00$\pm0.4$ & 3.69$\pm0.24$ \\
    10 & 1.4 & 11.2 & 0.02 & 0.2 & 1.945 & 59.57$\pm3.37$ & 4.63$\pm0.19$ & 1.64$\pm0.14$ & 3.37$\pm0.12$ \\
    11 & 1.4 & 11.2 & 0.02 & 0.3 & 1.945 & 57.95$\pm2.65$ & 4.35$\pm0.16$ & 1.36$\pm0.09$ & 3.08$\pm0.09$ \\
    12 & 1.4 & 11.2 & 0.02 & 0.4 & 1.945 & 55.12$\pm3.45$ & 4.12$\pm0.19$ & 0.98$\pm0.04$ & 2.78$\pm0.10$ \\
    13 & 1.4 & 11.601 & 0.02 & 0.35 & 2.0 & 52.51$\pm2.43$ & 4.47$\pm0.16$ & 1.16$\pm0.05$ & 2.90$\pm0.08$ \\ 
    14 & 1.4 & 11.601 & 0.02 & 0.35 & 2.5 & 52.93$\pm3.91$ & 4.33$\pm0.27$ & 1.14$\pm0.13$ & 2.26$\pm0.11$ \\ 
    15 & 1.4 & 11.601 & 0.02 & 0.35 & 3.0 & 55.04$\pm3.55$ & 4.19$\pm0.40$ & 1.13$\pm0.25$ & 1.90$\pm0.13$ \\ 
    16 & 1.4 & 11.601 & 0.02 & 0.35 & 4.0 & 55.28$\pm2.37$ & 4.04$\pm0.26$ & 1.11$\pm0.13$ & 1.38$\pm0.09$ \\
    17 & 1.4 & 11.601 & 0.02 & 0.1 & 2.5 & 57.92$\pm3.15$ & 4.87$\pm0.22$ & 1.36$\pm0.22$ & 2.79$\pm0.08$ \\
    18 & 1.4 & 11.601 & 0.02 & 0.2 & 2.5 & 55.99$\pm1.67$ & 4.57$\pm0.14$ & 1.24$\pm0.07$ & 2.53$\pm0.08$ \\
    19 & 1.4 & 11.601 & 0.02 & 0.3 & 2.5 & 54.50$\pm3.80$ & 4.36$\pm0.25$ & 1.17$\pm0.09$ & 2.35$\pm0.13$ \\
    20 & 1.4 & 11.601 & 0.02 & 0.4 & 2.5 & 53.35$\pm3.03$ & 4.06$\pm0.21$ & 1.05$\pm0.04$ & 2.14$\pm0.12$ \\
    
    \toprule
    Model & M & R & $\rm Z_{CNO}$ & $\rm Q_{e}$ & $\rm \dot{M}$  & $\rm t_{rise}$ & $\rm \tau$ & $\rm \tau_{e}$ & $\rm P_{ign}$ \\
    $\rm Number$ & $\rm M_{\odot}$ & $\rm km$ & $ $ & $\rm MeV/u$ & $ 10^{-9}~M_{\odot}/\rm yr$ & $\rm s$ & $\rm s$ & $\rm s$ & $\rm 10^{22}~dyn~cm^{-2}$ \\
    \hline
    1 & 1.3 & 11.2 & 0.01 & 0.1 & 1.945 & 6.08$\pm0.53$ & 22.37$\pm2.57$ & 40.34$\pm3.46$ & 2.24$\pm0.04$ \\
    2 & 1.5 & 11.2 & 0.01 & 0.1 & 1.945 & 5.41$\pm0.81$ & 15.26$\pm2.41$ & 28.43$\pm3.39$ & 2.69$\pm0.08$ \\
    3 & 1.7 & 11.2 & 0.01 & 0.1 & 1.945 & 4.42$\pm0.81$ & 10.97$\pm1.81$ & 18.91$\pm2.42$ & 3.46$\pm0.13$ \\
    4 & 1.8 & 11.2 & 0.01 & 0.1 & 1.945 & 4.55$\pm0.33$ & 8.93$\pm0.65$ & 15.50$\pm0.99$ & 3.93$\pm0.10$ \\
    5 & 1.4 & 11.2 & 0.01 & 0.1 & 1.945 & 5.62$\pm0.69$ & 16.84$\pm3.05$ & 31.53$\pm4.03$ & 2.51$\pm0.08$ \\
    6 & 1.4 & 12.0 & 0.01 & 0.1 & 1.945 & 6.32$\pm0.64$ & 22.57$\pm3.59$ & 42.55$\pm4.74$ & 2.12$\pm0.06$ \\
    7 & 1.4 & 12.5 & 0.01 & 0.1 & 1.945 & 6.09$\pm0.53$ & 25.34$\pm3.93$ & 47.07$\pm5.57$ & 1.89$\pm0.04$ \\
    8 & 1.4 & 13.0 & 0.01 & 0.1 & 1.945 & 6.07$\pm0.84$ & 25.88$\pm6.16$ & 47.90$\pm10.51$ & 1.75$\pm0.05$ \\
    9 & 1.4 & 11.2 & 0.02 & 0.1 & 1.945 & 5.27$\pm0.86$ & 14.54$\pm3.57$ & 24.91$\pm5.35$ & 2.30$\pm0.11$ \\
    10 & 1.4 & 11.2 & 0.02 & 0.2 & 1.945 & 5.44$\pm1.15$ & 15.93$\pm2.57$ & 28.73$\pm2.97$ & 2.12$\pm0.06$ \\
    11 & 1.4 & 11.2 & 0.02 & 0.3 & 1.945 & 5.68$\pm0.74$ & 17.65$\pm3.30$ & 32.12$\pm3.69$ & 1.98$\pm0.04$ \\
    12 & 1.4 & 11.2 & 0.02 & 0.4 & 1.945 & 6.61$\pm0.52$ & 19.06$\pm4.55$ & 36.82$\pm6.84$ & 1.49$\pm0.03$ \\
    13 & 1.4 & 11.601 & 0.02 & 0.35 & 2.0 & 6.16$\pm0.63$ & 17.61$\pm3.87$ & 33.12$\pm5.54$ & 1.75$\pm0.03$ \\
    14 & 1.4 & 11.601 & 0.02 & 0.35 & 2.5 & 6.51$\pm0.59$ & 17.83$\pm3.32$ & 32.70$\pm5.44$ & 1.84$\pm0.06$ \\
    15 & 1.4 & 11.601 & 0.02 & 0.35 & 3.0 & 6.33$\pm0.63$ & 18.96$\pm4.38$ & 35.73$\pm6.80$ & 1.90$\pm0.08$ \\
    16 & 1.4 & 11.601 & 0.02 & 0.35 & 4.0 & 6.48$\pm0.66$ & 18.13$\pm4.13$ & 33.32$\pm7.03$ & 2.03$\pm0.10$ \\
    17 & 1.4 & 11.601 & 0.02 & 0.1 & 2.5 & 6.44$\pm0.70$ & 19.16$\pm3.7$ & 36.04$\pm4.96$ & 2.08$\pm0.06$ \\
    18 & 1.4 & 11.601 & 0.02 & 0.2 & 2.5 & 6.32$\pm0.59$ & 18.82$\pm3.51$ & 34.95$\pm4.41$ & 1.95$\pm0.05$ \\
    19 & 1.4 & 11.601 & 0.02 & 0.3 & 2.5 & 6.42$\pm0.56$ & 19.35$\pm4.13$ & 35.29$\pm6.41$ & 1.88$\pm0.06$ \\
    20 & 1.4 & 11.601 & 0.02 & 0.4 & 2.5 & 6.40$\pm0.53$ & 16.95$\pm4.13$ & 31.88$\pm6.50$ & 1.77$\pm0.06$ \\
    \hline
\toprule
\end{tabular}
\label{tab:p}
\end{table}


\begin{deluxetable*}{cccccccccccc}

\tablenum{2}
\tablecaption{Physical quantities of burst models for HERES code.}

\tablewidth{0pt}
\tablehead{
\colhead{EOS} & \colhead{$ M$} & \colhead{$ R$} & \colhead{$\rm Z_{CNO}$} & {$ \dot{M}$} & {$\rm \alpha$} & {$\rm \tau$} & \colhead{$\rm \Delta t$} & \colhead{$\rm E_{burst}$} & \colhead{$\rm L_{peak}$} & \colhead{$\rm t_{rise}$} & \colhead{$\rm \tau_{e}$}  \\
\colhead{$ $} & \colhead{$ M_{\odot}$} & \colhead{$\rm km$} & \colhead{$ $} & \colhead{$ 10^{-9}~M_{\odot}/\rm yr$} & \colhead{$ $} & \colhead{$\rm s$} & \colhead{$\rm h$} & \colhead{$\rm 10^{39}~erg$} & \colhead{$\rm 10^{38}~erg/s$} & \colhead{$\rm s$} & \colhead{$\rm s$}  }
\startdata
Togashi & 1.4 & 11.601 & 0.02 & 2.0 & 57.28$\pm1.44$ & 27.89$\pm1.81$ & 3.04$\pm0.07$ & 4.28$\pm0.08$ & 1.54$\pm0.10$ & 5.15$\pm0.44$ & 19.55$\pm2.07$  \\
Togashi & 1.4 & 11.601 & 0.02 & 2.5 & 53.11$\pm1.77$ & 39.14$\pm2.06$ & 2.17$\pm0.06$ & 4.12$\pm0.09$ & 1.06$\pm0.07$ & 4.41$\pm0.50$ & 29.19$\pm1.89$ \\
Togashi & 1.4 & 11.601 & 0.02 & 3.0 & 52.73$\pm2.11$ & 42.22$\pm2.23$ & 1.76$\pm0.05$ & 4.05$\pm0.12$ & 0.96$\pm0.08$ & 4.09$\pm0.48$ & 30.79$\pm1.94$\\
Togashi & 1.4 & 11.601 & 0.02 & 4.0 & 52.10$\pm1.08$ & 46.39$\pm1.22$ & 1.27$\pm0.02$ & 3.93$\pm0.08$ & 0.85$\pm0.03$ & 4.03$\pm0.31$ & 34.00$\pm1.87$ \\
\enddata
\tablecomments{
(a) The base heat {\it calculated} from the luminoty value on the crust surface is $Q_b=0.35~{\rm MeV/u}$~\citep{Dohi2021ApJ...923...64D}.\\
(b) The data are different from \cite{Dohi2021ApJ...923...64D} in that the adopted reaction rates of ${}^{64}{\rm Ge}(p,\gamma){}^{65}{\rm As}$ and ${}^{65}{\rm As}(p,\gamma){}^{66}{\rm Se}$ are \cite{Cyburt2016ApJ...830...55C} for former, while \cite{2016ApJ...818...78L} for latter.
}
\end{deluxetable*}
\label{tab:Heres}

\begin{deluxetable*}{ccccccccccccccc}
	\tablenum{3}
	\tablecaption{Nuclear reaction network of rp.net (304 species) and APRX3 (88 species)} \label{tab:net}
	\tablewidth{0pt}
	\tablehead{
	 & \colhead{rp.net} &  \colhead{APRX3} &&   & \colhead{rp.net} &  \colhead{APRX3}  &&   & \colhead{rp.net} &  \colhead{APRX3}  &&  & \colhead{rp.net} &  \colhead{APRX3} \\
	\colhead{$Z=$1--13} & \multicolumn{2}{c}{$A$} && \colhead{$Z=$14--26} & \multicolumn{2}{c}{$A$}   &&    \colhead{$Z=$27--39} & \multicolumn{2}{c}{$A$} && \colhead{$Z=$40--52} & \multicolumn{2}{c}{$A$}} 
	\startdata
	H & 1--3 & 1 &&  Si & 24--30 & 24--25  &&    Co & 51--57 & -- && Zr & 78--83 & 80,84   \\
	He & 3,4 & 4 && P & 26--31 & --   &&    Ni & 52--56 & 54,56,60 && Nb & 81--85 & 88  \\
	Li & 7 & -- &&  S & 27-34 & 28--30  &&    Cu & 54--63 & -- && Mo & 82--86 & 84  \\
	Be & 7,8 & -- &&  Cl & 30--35 & --  &&    Zn & 55--66 & 60,64 && Tc & 85--88 & 92  \\
	B & 8,11 & -- &&  Ar & 31--38 & 33-34  &&    Ga & 59--67 & -- && Ru & 86--91 & 88,90,92  \\
	C & 9,11,12 & 12 &&  K & 35--39 & --  &&    Ge & 60--68 & 62-64,68 && Rh & 89-93 & 96   \\
	N & 12--15 & -- &&  Ca & 36--44 & 37--40  &&    As & 64--69 & -- && Pd & 90-94 & 92,94,96,98   \\
	O & 13--18 & 14--16 &&  Sc & 39--45 & --   &&    Se & 65--72 & 68,72 && Ag & 94--98 & 97--98,102  \\
	F & 17--19 & -- &&  Ti & 40--47 & 42  &&    Br & 68--73 & -- && Cd & 95--99 & 102--106  \\
	Ne & 18--21 & 18 &&  V & 43--49 & --  &&    Kr & 69--74 & 72,76 && In & 98--104 & 99,102--107,109  \\
	Na & 20--23 & -- &&  Cr & 44--52 & 46  &&    Rb & 73--77 & -- && Sn & 99--105 & 100--109,112  \\
	Mg & 21--25 & 21--22 &&  Mn & 47--53 & --  &&    Sr & 74--78 & 76,80 && Sb & 106 & 106--108  \\
	Al & 22--27 & -- &&  Fe & 48--56 & 48,50  &&    Y & 77--82 & -- &&  Te & 107 & 107--109 \\
	\enddata
	\tablecomments{Nuclear reaction grids from \citet{Fisker2007astro.ph..3311L} and \citet{Dohi2020PTEP.2020c3E02D}, respectively}
    \end{deluxetable*}

\section{Correcting the quantities from local frame to the frame of a distant observer} \label{sec:red}
To compare the burst parameters with observations, it is crucial to correct the GR quantities from local reference frame of the NS surface to the frame of a distant observer, which we mark with the superscript ``$\infty$".

The timescale will be redshifted by 
\begin{equation} \label{eq:rt}
t^{\infty}=(1+z_g)t
\end{equation}
where $1+z_g=\frac{1}{\sqrt{1-\frac{2GM}{Rc^2}}}$.

The redshifted luminosity can be written as
\begin{equation} \label{eq:rl}
    L^{\infty}=\frac{L}{(1+z_g)^2}
\end{equation}

Because the burst energy $E_{\rm b}$ is obtained by integrating over the time (see equation~(\ref{eq:eb})), from equations~(\ref{eq:rt})(\ref{eq:rl}), the redshifted burst energy is given by
\begin{equation}
    E_{\rm b}^{\infty}=\frac{E_{\rm b}}{1+z_{\rm g}}
\end{equation}

Similarly, the redshifted mass accretion rate is given by

\begin{equation}
    \dot{M}^{\infty}=\frac{\dot{M}}{1+z_g}
\end{equation}

The local burst strength from equation~(\ref{eq:alpha}) can be redshifted by 
\begin{equation}
    \alpha^{\infty}=\frac{\alpha}{1+z_g}
\end{equation}

One can transfer the time scales (e.g. recurrence time $\Delta t$, rise time $t_{\rm rise}$, duration time $\tau$, e-folding time $\tau_{\rm e}$ ), the luminosities ( e.g. peak luminosity $L_{\rm pk}$), the burst energy $E_{\rm b}$, mass accretion rate $\dot{M}$ and the burst strength $\alpha$ from the local reference frame to an observer frame by the above formulas.

\end{document}